\documentclass[amssymb,amsfonts]{amsart}
\usepackage{amssymb, latexsym}

\newtheorem{thm}{Theorem}
\newtheorem{defn}[thm]{Definition}
\newtheorem{prop}[thm]{Proposition}

\newtheorem{cor}[thm]{Corallary}
\newtheorem{lem}[thm]{Lemma}
\newtheorem{rmk}[thm]{Remark}

\begin{document}
\title[D.E.C.~for generalized wave-maps]{Dominant energy condition and causality for
Skyrme-like generalizations of the wave-map equation}
\author[W. W. Wong]{Willie Wai-Yeung Wong}
\address{Department of Pure Mathematics and Mathematical Statistics\\
University of Cambridge\\ Cambridge, UK}
\email{ww278@dpmms.cam.ac.uk}
\thanks{The author is supported by the Commission of the European
Communities, ERC Grant Agreement No 208007.}

\begin{abstract}
It is shown in this note that a class of Lagrangian field theories
closely related to the wave-map equation and the Skyrme model obeys
the dominant energy condition, and hence by Hawking's theorem
satisfies finite speed of propagation. The subject matter is a
generalization of a recent result of Gibbons. 
\end{abstract}

\maketitle

\section{Introduction}

Recently Gibbons showed \cite{PDE:skyrm:Gibbons03} that the Skyrme
model obeys the dominant energy condition, and thus settling the
problem of causality for that equation. In this note we will give a
different proof of the same fact that easily generalizes to a class of
Lagrangian field theories that includes, as special cases, the
wave-map equation, the Skyrme model, and the Born-Infeld model. 

Let $(M,g)$ be an $m+1$ dimensional Lorentzian manifold, where sign
convention is taken to be $(-,+,+,\cdots)$, and let
$(N,h)$ be an $n$ dimensional Riemannian manifold. Let $\phi:M\to N$
be a $C^1$ map. Then the action of $\phi$ can be used to pull back the
metric $h$ onto $M$ as a positive semi-definite quadratic form on $TM$,
we write it as 
\[ \phi^*h(X,Y) = h(d\phi\cdot X, d\phi\cdot y) \]
where the left hand side is evaluated at a point $p\in M$ and the
right hand side at the point $\phi(p)\in N$ for $X,Y\in T_p M$.
Composing with the inverse metric $g^{-1}$ we obtain the so-called
\emph{strain tensor} $D^\phi$, a section of $T^1_1M$:
\begin{equation}\label{eq:def:strain}
D^\phi = g^{-1}\circ \phi^*h~,
\end{equation}
thus at every point $p$, $D^\phi$ is a linear transformation of
$T_pM$. Now, if $g$ were a Riemannian metric, then for a fixed basis
of $T_pM$, the matrix $(D^\phi)$ is positive semi-definite. This is,
unfortunately, no longer true in the Lorentzian case, and thus the
eigenvalues of $(D^\phi)$ are in general complex. 

Let $\{\lambda_1,\ldots,\lambda_k\}$ denote the non-zero eigenvalues,
counted with multiplicity, of $(D^\phi)$. Note that by elementary
linear algebra, using that $g$ is non-degenerate and $h$ is positive
definite, one easily sees that 
\begin{equation}\label{eq:kupperbound}
 k \leq \mathop{rank}(d\phi) \leq \min(m+1,n)~. 
\end{equation}

Recall the elementary symmetric polynomials
$s_j(\{\lambda_1,\ldots,\lambda_k\})$ given by 
\begin{equation}
s_j(\{\lambda_1,\ldots,\lambda_k\}) = \sum_{1\leq \alpha_1 < \alpha_2
< \cdots < \alpha_j \leq k} \prod_{i = 1}^{j}\lambda_{\alpha_i}
\end{equation}
with $s_0 = 1$ and $s_j = 0$ for all $j > k$. Observe that for the
$(m+1)\times (m+1)$ matrix $(D^\phi)$, the elementary symmetric polynomials 
correspond to the coefficients of the characteristic polynomial, and 
specifically $s_1 = \mathop{tr} (D^\phi)$ and $s_{m+1} = \det
(D^\phi)$. By abuse of notation, we will write $s_j(D^\phi)$ when we
mean the symmetric polynomials on the eigenvalues of $(D^\phi)$. Note
that $s_j(D^\phi)$ is independent of a basis chosen for the vector
space $T_pM$. 

For a given class $\mathcal{A}$ of maps $\phi:M\to N$, we write
\[
U_\mathcal{A}:= \{ v\in\mathbb{R}^{m+1}~ |~ v =
(s_1,\ldots,s_{m+1})(D^\phi)~,~\phi\in \mathcal{A}\}~. \]
\begin{defn}
For a given class $\mathcal{A}$, let  
$\mathcal{U}_\mathcal{A}\subset \mathbb{R}^{m+1}$ be an open set that contains
$U_\mathcal{A}\cup \{0\}$. An \emph{admissible function} 
$F:\mathcal{U}_\mathcal{A}\to\mathbb{R}$ 
for the class $\mathcal{A}$ is a sub-additive, concave function, that is
$C^1$ on the interior of $\mathcal{U}_\mathcal{A}$ and
continuous up to the boundary. 
\end{defn}
\begin{rmk}
In the definitions above, it only suffices to include terms up to $s_{m+1}$
in view of \eqref{eq:kupperbound}. Also, observe that sub-additivity
and concavity of $F$ immediately implies that $F(0) \geq 0$. 
\end{rmk}

\begin{defn}\label{defn:genwavemaps}
A Lagrangian field theory for the class $\mathcal{A}$ of maps 
$\phi:M\to N$ is said to be a
\emph{generalized wave-map}\footnote{For the lack of a better name.
Suggestions are welcome.} if the Lagrangian 
\[ L = F(s_1(D^\phi), s_2(D^\phi),\ldots,s_{m+1}(D^\phi)) \]
for an admissible $F$. Furthermore, we say that the generalized
wave-map is \emph{defocusing} if the first partial derivatives of $F$
are all non-negative, i.e.~$\partial_iF(v) \geq 0$ $\forall i =
1,\ldots,m+1$ and $\forall v\in\mathcal{U}_\mathcal{A}$. The generalized
wave-map is said to be \emph{zeroed} if $F(0) = 0$. Also, we shall 
refer to a generalized wave-map for which
$\partial_1F$ is non-vanishing as \emph{non-degenerate}. 
\end{defn}

The author hopes that the reason behind the nomenclature will be
evident after the proof of the dominant energy condition is developed.
We first give some examples of generalized wave-maps:
\begin{itemize}
\item 
Observe that if $L$ is a linear combination of the symmetric
polynomials $L = \sum c_is_i(D^\phi)$, than it is automatically a 
zeroed generalized wave-map.
If in addition the coefficients $c_i$ are all non-negative, then $L$ is
defocusing. In this case if $c_1 > 0$ then $L$ is non-degenerate. 
\item 
Take $(M,g)$ to be a static space-time, i.e.~$M =
\mathbb{R}\times\Sigma$ and $g = -\rho dt^2 \oplus \gamma$ where
$\rho$ is a positive function on $\Sigma$ and $\gamma$ is a Riemannian
metric on $\Sigma$. A static solution to the generalized wave-map is
one for which $\nabla_t\phi = 0$. The static solution for $L = s_1$ 
gives rise to the harmonic map equation from $\Sigma\to N$, while for 
the case $n > m$, $L = \sqrt{s_m}$ (recall that $\dim M = m+1$), the equation 
becomes the minimal surface equation for the embedding of $\Sigma$ into
$N$. For the minimal surface equation we take $\mathcal{U}_\mathcal{A}
= \bar{\mathbb{R}}^{m+1}_+$. 
\item
In the Lorentzian case, $L=s_1$ is simply the wave-map equation. For
$L = c_1 s_1 + c_2 s_2$ where $c_1, c_2 > 0$ are coupling constants,
we recover the original Skyrme model if we take $(N,h)$ to be $SU(2)$
with the bi-invariant metric. In particular, the Skyrme model is a
defocusing, zeroed, non-degenerate, generalized wave-map in the terminology
adopted in the present paper. 
\item
Let $b > 0$ be a fixed large constant. We can restrict $\phi$ to only
consider those maps such that the real parts of the eigenvalues of
$D^\phi$ are greater than $-b$. Then letting 
\[ F = \sqrt{\det (b\cdot Id + D^\phi)} - \sqrt{\det(b\cdot Id)} \]
defined on $\mathcal{U}_\mathcal{A}$ being the set where $\det (b\cdot
Id + D^\phi)\geq 0$, 
we get the zeroed, defocusing, non-degenerate, generalized wave-map
also known as the Born-Infeld model. 
\end{itemize}

Before stating the main theorem, we recall the statement of the
dominant energy condition. Recall that the (covariant) stress-energy 
tensor $T\in\Gamma(T^0_2M)$ for a Lagrangian field theory is given by a
variational derivative for the Lagrangian \emph{density} relative to
the inverse metric, 
\begin{equation}\label{eq:def:setensor}  T\sqrt{|\det g|} := \frac{\delta [L \sqrt{|\det g|}]}{\delta
g^{-1}} = \left(\frac{\delta L}{\delta g^{-1}}- \frac12 L g
\right)\sqrt{|\det g|}~.
\end{equation}

\begin{defn}
The stress-energy tensor $T$ is said to obey the \emph{dominant energy
condition} at a point $p\in M$ if $\forall X\in T_pM$ such that
$g(X,X) < 0$, the following two conditions are satisfied
\begin{subequations}
\begin{align}
\label{eq:DECpastpointing} T(X,X) & > 0 \\
\label{eq:DECcausal} [ T\circ g^{-1} \circ T ](X,X) &\leq 0
\end{align}
\end{subequations}
unless $T$ vanishes identically.
\end{defn}
\begin{rmk}
The definition is equivalent to the classical statements (see,
e.g.~section 4.3 in \cite{GR:genbook:HawkElli} or chapter 9 of
\cite{GR:genbook:Wald}) of the dominant energy condition. 
Observe that \eqref{eq:DECcausal} gives that the vector $g^{-1}\circ
T\circ X$ is a causal vector for any time-like vector $X$, and
\eqref{eq:DECpastpointing} gives that the vector $g^{-1}\circ
T\circ X$ has opposite time-orientation as the time-like vector $X$. 
\end{rmk}

Now we state the main theorem 
\begin{thm}\label{thm:main}
A defocusing generalized wave-map obeys the dominant
energy condition. 
\end{thm}
First we claim that it would suffice to prove the theorem for each
$s_i$. The fellowing lemma is a general statement on a convexity
property of Lagrangian field theories. 
\begin{lem}
Let $F$ be a sub-additive, concave function as in
Definition \ref{defn:genwavemaps}. Let $T_i$ denote the stress-energy
tensor corresponding to the Lagrangian $L_i$. Assume
that $T_i$ obeys the dominant energy condition, or, equivalently,
the vectors $Y_i = g^{-1}\circ T_i\circ X$ are all past-causal for any
fixed future time-like $X$. Then $L = F(L_1,\ldots,L_{m+1})$ also
obeys the dominant energy condition if $L$ is defocusing. 
\end{lem}
\begin{proof}
The stress-energy tensor $T$ can be written, using
\eqref{eq:def:setensor}, as 
\[ T = \sum_{i = 1}^{m+1} \partial_i F \cdot \frac{\delta
L_i}{\delta g^{-1}} - \frac12 F g = \sum_{i=1}^{m+1}
\partial_i F \cdot T_i - \frac12 (F - \sum_{i=1}^{m+1}\partial_i
F\cdot L_i) g~.\]
Now considering $g^{-1}\circ T\circ X$, the first term in the above
expression contributes $\sum \partial_i F \cdot Y_i$. Since $L$ is
defocusing, this is a positive linear combination of past-causal
vectors, and hence by elementary Minkowskian geometry, is still
past-causal. For the second term, since $g^{-1}\circ g\circ X = X$, to
show that it is also past-causal it suffices to show that
\[ F \geq \sum_{i=1}^{m+1}\partial_i F\cdot L_i~. \]
But this follows from the fact that $F$ is concave and $F(0) \geq 0$. 
\end{proof}

Unfortunately, it is immediately clear that the theorem may not be
strong enough in certain cases for practical application. This is 
because the vanishing of $T$
does not guarantee that the map $\phi$ is trivial. 
For example, using that $s_j = 0$ if $j > \mathop{rank}(d\phi)$, it is
immediate that if locally around the point $p$, $\phi$ is
one-dimensional, then for any metric $g$, $s_j(D^\phi) = 0$ if $j \geq
2$. On the other hand, this failure of the dominant energy condition
arises from a degeneracy which forces the stress-energy tensor to
be a null stress tensor in the language of Christodoulou
\cite{PDE:genbook:Chris00}, which we can ``normalize'' away by taking
$L$ to be zeroed. We claim that this is the only possible failure. 
\begin{prop}\label{prop:rankcondition}
For $L = s_i$, $T$ obeys the dominant energy condition. Furthermore,
$T = 0$ at a point $p$ if and only if $i > \mathop{rank}(d\phi |_p)$. 
\end{prop}

From this proposition one immediately sees the following energy bound
for smooth solutions of the generalized wave-map equation. 
\begin{cor}
If $\phi$ is the solution to a defocusing, non-degenerate, zeroed, 
generalized
wave-map, and if $T = 0$ on a connected open domain $\mathcal{B}$ of 
$M$, then $\phi$ is constant on $\mathcal{B}$. 
\end{cor}
By applying Hawking's energy conservation theorem (see section 4.3 in
\cite{GR:genbook:HawkElli}) the above corollary implies that
defocusing, non-degenerate, zeroed, generalized wave-maps have finite speed of
propagation (also known as the domain of dependence condition). 

In principle, if one has advanced knowledge on a lower bound to the 
rank of the map $\phi$, one can also obtain analogous statements for
degenerate cases. We leave such trivial generalizations to the reader. 

The author would like to thank Nick Manton and Claude Warnick for
introducing him to the problem, and to Mihalis Dafermos for useful
discussions. 

\section{A formula for the stress-energy tensor and proof of the main
proposition}

In this section, we'll derive first derive a formula for the 
stress-energy tensor. We will begin by making a geometric
observation and obtain, almost immediately, a simple tensorial formula
for the Lagrangian. Taking the formal variational derivative of 
the Lagrangian leads to a
tensorial expression for the stress-energy tensor, from which
Proposition \ref{prop:rankcondition} follows via simple linear
algebra. 

Consider a real vector space $V$. Let $A$ be a linear transformation on
$V$. Then $A$ naturally extends to a linear transformation, which we
denote $A^{\sharp j}$, on
$\Lambda^j(V)$, the space of alternating $j$-vectors over $V$. A bit
of basic linear algebra (perhaps by extending $V$ to
$V\otimes_{\mathbb{R}}\mathbb{C}$ and taking a basis of eigenvectors)
shows that $s_j(A)$ is proportional to
$\mathop{tr}_{\Lambda^j(V)}A^{\sharp j}$. Now, letting $V = T_pM$ and
$A = D^\phi = g^{-1}\circ \phi^*h$, we observe that 
\[ (D^\phi)^{\sharp j} = (g^{-1})^{\sharp j} \circ \phi^* (h^{\sharp
j})~, \]
or, to put it in words, $(D^\phi)^{\sharp j}$ is obtained from first
taking the induced metric $h^{\sharp j}$ on alternating $j$-vectors in
$T_{\phi(p)}N$, pulling it back via $\phi$, and composing it with the
induced metric $(g^{-1})^{\sharp j}$ for the alternating $j$-forms. In
index notation, this can be written as
\[
[(D^\phi)^{\sharp j}]^{b_1\ldots b_j}_{a_1\ldots a_j} =
g^{b_1c_1}\cdots g^{b_jc_j} (\phi^*h)_{a_1 [c_1|} 
(\phi^*h)_{a_2 |c_2|}\cdots (\phi^*h)_{a_{j-1} |c_{j-1}|}
(\phi^*h)_{a_j |c_j]}
\]
where the bracket notation in the indices denotes full
anti-symmetrization of the $\{c_1,\ldots,c_j\}$ indices. For a
Lagrangian proportional to an $s_j$,  we can assume 
\begin{equation}\label{eq:lagrangiantensorly}
L = [(D^\phi)^{\sharp j}]^{b_1\ldots b_j}_{a_1\ldots a_j} =
g^{a_1[c_1|}\cdots g^{a_j|c_j]} 
(\phi^*h)_{a_1 c_1} \cdots (\phi^*h)_{a_jc_j}~.
\end{equation}
It is simple to check, using $(D^\phi) = \mathop{diag}(-1,1,1,\ldots)$
that the above expression has the correct sign: that $L$ defined thus
is a positive multiple of $s_j$. 

One can also arrive at \eqref{eq:lagrangiantensorly} purely from a
linear algebra point of view. Let $p_j$ be the power sum
\[ p_j(\{\lambda_1,\ldots,\lambda_k\}) = \sum_{i=1}^{k} \lambda_i^j~. \]
Recall that we have Newton's identity
\[ j\cdot s_j = \sum_{i=1}^j (-1)^{i-1}e_{j-i}p_i \]
which allows us to express $s_j$ as a rational polynomial in $p_i$'s.
Now, by definition, it is clear that
\[ p_j(D^\phi) = \mathop{tr} [(D^\phi)^j] \]
where $(D^\phi)^j$ is the $j$-fold composition of $D^\phi$. It is easy
to check then, for some $E$
\[ s_j = g^{a_1b_1}\cdots g^{a_jb_j} E_{b_1\ldots b_j}^{c_1\ldots c_j}
(\phi^*h)_{a_1c_1}\cdots (\phi^*h)_{a_jc_j}~. \]
Newton's identity reduces to a generating condition for $E$ based on
the Kronecker $\delta$ symbols,
\begin{align*}
E_b^c &= \delta_b^c~,\\
j E_{b_1\ldots b_j}^{c_1\ldots c_j} &= \sum_{i = 1}^j
(-1)^{i-1}E_{b_1\ldots b_{j-i}}^{c_1\ldots c_{j-i}}
\delta_{b_{j-i+1}}^{c_{j-i}}\delta_{b_{j-i+2}}^{c_{j-i+1}}\cdots\delta_{b_j}^{c_{j-i+1}}~.
\end{align*}
A direct computation which we omit here shows that then in fact the 
invariant $E_{b_1\ldots b_j}^{c_1\ldots c_j}$ is a positive rational 
multiple of the generalized Kronecker symbol 
$\delta_{b_1\ldots b_k}^{c_1\ldots c_j}$, from which we recover
\eqref{eq:lagrangiantensorly}. 

Now, the object we are interested in, given a time-like vector $X$, is
the one-form $T(X,\cdot)$. Since $T$ is tensorial, we can assume $X$
has unit length. Fix some $j$, let the Lagrangian be proportional to
$s_j$ as given by \eqref{eq:lagrangiantensorly}. By the symmetry
property, we can write $T(X,\cdot)$ in index notation:
\begin{equation}\label{eq:setensortensorly}
T_{ab}X^b = j X^{[b|}g^{a_2|c_2|}\cdots g^{a_j|c_j]}(\phi^*h)_{a
b} \cdots (\phi^*h)_{a_jc_j} - \frac12 g_{ab}X^b L
\end{equation}

\begin{proof}[Proof of Proposition \ref{prop:rankcondition}]
Consider a orthonormal basis for $T_pM$ relative to $g$. Since we
assumed $X$ unit, let $e_0 = X$ and $\{e_i\}_{1\leq i\leq m}$ are all
space-like. We can take $j \leq m+1$ as otherwise $T$ is identically
0. Then we notice that a basis for $\Lambda^j(T_pM)$ is given by 
\[ \{ e_0 \wedge e_{\alpha_1}\wedge \cdots \wedge e_{\alpha_{j-1}}
\}_{1\leq \alpha_1 < \cdots < \alpha_{j-1}\leq m} \cup \{
e_{\alpha_1}\wedge \cdots \wedge e_{\alpha_j}\}_{1\leq \alpha_1 <
\cdots < \alpha_j\leq m}~. \]
We write the first set as $\Lambda^j_\perp$ and the second set as
$\Lambda_\parallel^j$. Using the normalization that $v\wedge w =
v\otimes w - w\otimes v$, we find that each of the element in
$\Lambda^j_\perp$ has norm $-j!$ while the elements in
$\Lambda^j_\parallel$ has norm $j!$. 

To show that $T(X,X) > 0$ generically, we observe that under the
expansion \eqref{eq:setensortensorly}, the first term corresponds to 
\[ \sum_{\omega\in \Lambda_\perp^j} \phi^* ( h^{\sharp j}
)(\omega,\omega)~, \]
while the second term corresponds to 
\[ \frac{1}{2} \left( - \sum_{\omega\in\Lambda_\perp^j} \phi^* (
h^{\sharp j})(\omega,\omega)  +
\sum_{\omega\in\Lambda_\parallel^j}
\phi^* (h^{\sharp j})(\omega,\omega)\right)~. \]
So summing them gives 
\[ \frac{1}{2} \left( \sum_{\omega\in\Lambda_\perp^j} \phi^* (
h^{\sharp j})(\omega,\omega) +
\sum_{\omega\in\Lambda_\parallel^j}
\phi^* (h^{\sharp j})(\omega,\omega)\right) \]
which is non-negative by the fact that $\phi^* (h^{\sharp j})$ is a
positive semi-definite quadratic form on $\Lambda^j(T_pM)$.
Furthermore, observe that since
$\Lambda^j_\parallel\cup\Lambda^j_\perp$ is a basis, its push-forward
$\phi_*\Lambda^j_\parallel\cup \phi_*\Lambda^j_\perp$ spans
$\Lambda^j(\phi_* T_p^M) \subset \Lambda^j(T_{\phi(p)}N)$. Thus by the
fact that $h$ (and hence the induced metric $h^{\sharp j}$) is
positive definite, we conclude that $\Lambda^j(\phi_* T_p^M) = \{
0\}$, which proves the assertion that $T$ vanishes only when $j >
\mathop{rank}(d\phi)$. 

To show \eqref{eq:DECcausal}, we observe that
\[
X^aT_{ac}g^{cd}T_{db}X^b = - T(X,X)^2 + \sum_{i = 1}^m T(X,e_i)^2~.
\]
The first thing to note is that $T(X,e_i)$ does not have any 
contribution from the second term in \eqref{eq:setensortensorly}. For
the first term, a quick computation shows that $T(X,ei)$ corresponds
to 
\[ \sum_{\eta \in \Lambda^{j-1}_\parallel} \phi^*(h^{\sharp j})(e_0\wedge \eta,
e_i\wedge\eta) \]
so 
\begin{align*}
|\sum_{i=1}^m T(X,e_i)^2 | & \leq ( \sum |T(X,e_i)| )^2 \\
& \leq ( \sum_{i = 1}^m \sum_{\eta\in \Lambda^{j-1}_\parallel}
|\phi^*(h^{\sharp j})(e_0\wedge \eta, e_i\wedge\eta)|)^2 \\
& \leq \frac14(\sum_{\eta\in \Lambda^{j-1}_\parallel}\phi^*(h^{\sharp
j})(e_0\wedge \eta,e_0\wedge \eta) + \sum_{i=1}^m\phi^*(h^{\sharp
j})(e_i\wedge \eta,e_i\wedge \eta))^2 \\
& = \frac14(\sum_{\eta\in \Lambda^{j-1}_\parallel}\sum_{i=0}^m\phi^*(h^{\sharp
j})(e_i\wedge \eta,e_i\wedge \eta))^2 \\
& = T(X,X)^2
\end{align*}
And therefore \eqref{eq:DECcausal} is satisfied. 
\end{proof}

\bibliographystyle{amsalpha}
\bibliography{../bib_files/master.bib}

\end{document}